\documentclass[12pt]{article}
\usepackage{epsfig}

\newcommand{\be}{\begin{eqnarray}}
\newcommand{\ee}{\end{eqnarray}}
\newcommand{\sll}{\raise.15ex\hbox{$/$}\kern-.43em\hbox{$l$}}
\newcommand{\slp}{\raise.15ex\hbox{$/$}\kern-.43em\hbox{$p$}}
\newcommand{\slq}{\raise.15ex\hbox{$/$}\kern-.43em\hbox{$q$}}
\newcommand{\slk}{\raise.15ex\hbox{$/$}\kern-.43em\hbox{$k$}}
\newcommand{\slepsilon}{\raise.15ex\hbox{$/$}\kern-.53em\hbox{$\epsilon$}}
\newcommand{\gsim}{\mbox{\raisebox{-0.6ex}{$\stackrel{>}{\sim}$}}\:}
\newcommand{\lsim}{\mbox{\raisebox{-0.6ex}{$\stackrel{<}{\sim}$}}\:}

\begin{document}

\bibliographystyle{unsrt}
\footskip 1.0cm

\thispagestyle{empty}

\begin{center}{\Large\bf{Proton and anti-proton production
in the forward region of d+Au collisions 
at RHIC from the color glass condensate}}\\

\vspace{1in}
{\large  Arata Hayashigaki}\\

\vspace{.2in}
{\it 
Institut f\"ur Theoretische Physik, J.~W.~Goethe Universit\"at\\
Max-von-Laue Strasse 1\\
D-60438 Frankfurt am Main, Germany}

\end{center}

\vspace*{25mm}

\begin{abstract}

\noindent
The power-law tail of high-$p_t$ $\pi^\pm$ spectra observed in forward
d+Au collisions at RHIC can be attributed to the power-law decrease of
the dipole forward scattering amplitude appearing in the color glass
condensate (CGC) approach.  Forward particle production probes the
small-$x$ gluon distribution of the target nucleus where its anomalous
dimension is rather flat ($\gamma= 0.6\sim 0.8$) for moderately high
$p_t$ ($\lsim 5$ GeV), and where the leading-twist DGLAP approximation
is not valid.  In the same framework, we examine $p$ and $\bar{p}$
production using baryon fragmentation functions parameterized in the
Lund fragmentation scheme.  This provides a good description of the
forward $\bar{p}$ spectrum while it underestimates the $p$ data by as
much as a factor of $2\sim 3$ at $p_t\lsim 4$ GeV. Part of this
anomalous baryon excess can be attributed to surviving constituent
diquarks from the deuteron projectile. Thus, the contribution from
diquark scattering may play an essential role for forward baryon
formation.
\end{abstract}
\newpage

\section{Introduction}
\label{sec_intro}

One mysterious problem in high-energy p(d)+A collisions
is a baryon excess seen in the $p/\pi^+$ ratio,
observed at semi-hard transverse momenta ($p_t\gsim 1$ GeV)
over a wide range of rapidity. Measurements include the central
rapidity region
at Fermilab ($\sqrt{s}=27.4,38.8$ GeV)~\cite{Fermilab} and
RHIC ($\sqrt{s}=200$ GeV)~\cite{STAR_Adams} as well as
forward rapidities ($y\sim 3$) at RHIC~\cite{BRAHMS_Debbe}.
In a phenomenological picture,
the standard string models assume a diquark-quark string structure
of the proton beam and dominance of soft processes
for leading baryon production 
as in the inside-outside cascade picture.
This leads to baryon production
only near beam rapidity and with small $p_t$.
To explain the baryon production near midrapidity
two additional nonperturbative 
models have been proposed: the diquark breaking~\cite{Boris} 
and gluon junction~\cite{Kharzeev} mechanisms. However,
these models are also restricted to small momentum
transfer ($p_t\lsim 2$ GeV).

Next-to-leading order (NLO) perturbative QCD calculations with standard
fragmentation functions cannot be reconciled with the forward-rapidity
data from d+Au collisions at RHIC, either. The data clearly show a
large baryon excess~\cite{GSV}.  Meanwhile, a pointlike diquark
picture involving the diquark form-factor gives a good description of
the FNAL data near midrapidity over $2\lsim p_t\lsim 10$
GeV~\cite{diquark_pict}.  Thus, the study of high-$p_t$ baryon spectra
in high-energy p(d)+A collisions is very challenging and provides an
opportunity for revealing soft successive processes from proton
breakup to their parton fragmentations.

It is also known that physics toward forward rapidity of the incident
proton from midrapidity is accompanied by a change of particle
production mechanism: transition from string breaking region, where
the particle production is dominated by quark-antiquark pairs created
from vacuum, to projectile fragmentation region, where initial charge
and isospin of the projectile are most likely conserved for the
particle production.  A signal of such a transition appears, for
instance, in antiparticle-to-particle ratios from p+p collisions at
large transverse momentum, which are independent of rapidity at less
than $y\sim 1.5$ but above this point decrease with rapidity,
irrespective of the detected hadron species~\cite{BRAHMS_pp}.  In the
forward-rapidity region, therefore, diquarks carrying away
a large fraction ($\sim 2/3$) of the incoming proton momentum will
increase the importance of its role for leading baryon production in the
projectile fragmentation region~\cite{diquark}.

In very high-energy p+A collisions, in particular for the kinematics
realized at forward rapidity, a dense target close to the black-body
(or unitarity) limit should destroy completely the coherence of the
projectile partons since all of them experience large transverse kicks
on the order of the saturation scale (onset of gluon saturation
phenomenon) and hence fragment independently, mainly into
pions~\cite{DGS}.  This leads to a strong suppression of the forward
baryon number.  Also, according to this idea the $p_t$ spectrum is
expected to be rather flat up to the saturation scale.  Although
recent BRAHMS d+Au data at $y\sim 3$~\cite{BRAHMS_Debbe} are almost
flat over the range $1<p_t\lsim 3$ GeV, these data do not seem to
support this idea of forward baryon number suppression at RHIC energy.
Rather, the ratio $p/\pi^+\sim 0.8$ is substantially larger than for
p+p collisions~\footnote{Pythia simulations~\cite{Pythia} also
underpredict this ratio by a factor $2\sim 3$ as compared to the
data~\cite{BRAHMS_Debbe}.}  in the range $1<p_t\lsim 3$ GeV.
Therefore, additional quantitative theoretical studies of forward
baryon production in
p(d)+A collisions, and its relation to saturation physics is of
importance in detailed comparisons with the data.

Another important experimental observation is that the ratio of hadron
$p_t$-distributions in d+Au versus p+p collisions is suppressed at
large rapidity~\cite{Arsene:2004ux}, while showing a
(species-dependent) Cronin enhancement at
midrapidity~\cite{rhic,AccGy}.  Within the Color Glass Condensate
(CGC) formalism~\cite{nonlin}, the disappearance of the Cronin peak at
moderate $p_t$ is induced by quantum evolution of the dense coherent
gluons in the target~\cite{forsup,dima,jjm}, therefore signaling the
emergence of saturation physics\footnote{Other scenarios also were
proposed to explain the phenomenon~\cite{rudi,qiu-vitev}.}.  
In the CGC picture,
the overall growth of the gluon density is decelerated with rapidity
because the rate of gluon fusion becomes comparable with that of gluon
emission.  With increasing rapidity, thus the cross section of the
gold target at fixed impact parameter grows less rapidly than that of
a proton target~\cite{JK}.

Along the lines of the CGC formalism, Ref.~\cite{aaj} developed an
asymmetric ${\rm DGLAP_{proj}}$~\cite{dglap} $\otimes$
${\rm BFKL_{targ}}$~\cite{bfkl} factorization
scheme which accounts for recoil effects due to collinear gluon
radiation in the projectile, because radiation becomes important with
higher $p_t$~\cite{kovmuel}.  Furthermore, in Ref.~\cite{aaj2} a
parameterization of the anomalous dimension $\gamma$ describing the
quantum $x$-evolution away from the Glauber-Mueller~\cite{Muell} or
McLerran-Venugopalan saturation models~\cite{MV} (which do not include
evolution due to fixed $\gamma=1$) over a wide range of rapidity
was provided.  Indeed, these formulations give a good description of
charged hadron or neutral pion production.  Then, the importance of
recoil effects and the appropriate use of the $p_t$ and $y$-dependent
anomalous dimension in the high $p_t$ region were discussed
quantitatively~\cite{aaj,aaj2}.

In this paper, we discuss $\pi^\pm$, $p$, $\bar{p}$ production with
high transverse momentum ($p_t\gsim 1$ GeV) in the deuteron
fragmentation region ($y=3.0$) of d+Au collisions at RHIC energy.  At
such rapidity the light-cone momentum fractions of the partons
participating in the scattering becomes very asymmetric between
projectile (d) and target (Au)~\cite{adjam}; large-$x$ quarks with
$x_p=O(10^{-1})$ from the deuteron collide with (many) small-$x$
gluons, $x_A=O(10^{-3}\sim 10^{-4})$ from the gold target (see figures
in~\cite{aaj,aaj2}).  Therefore, one should treat the deuteron as a
dilute and the gold nucleus as a dense object involving saturation
effects. Its saturation scale is given by $Q_s^2\sim A^{1/3}e^{\lambda
y}$ with $A\simeq200$ the mass number and $y$ the rapidity of the
target gluons. $\lambda$ is a constant describing the growth rate of
the saturation momentum with $1/x$.  Thus, the saturation effects in
the target are most evident at large $y$ and the black-body limit of
the target discussed in \cite{DGS} is most likely expected there.  On
the other hand, hadron production close to the deuteron beam rapidity
is dominated by its valence quarks (or their diquark bound state).
Energy loss of these leading quarks traversing the cold nucleus might
occur beyond the leading $\log Q^2$ approximation~\cite{FS}.

We will basically use the same expressions for the single-inclusive
hadron production cross sections as discussed in Refs.~\cite{aaj,aaj2}
in this paper, but will reformulate them slightly to include mass
corrections (massless particles have been assumed in
Refs.~\cite{aaj,aaj2}); cf.\ Sec.~\ref{sec2}.  The collinearly
factorized form derived here represents our basic equation. In
subsequent sections we discuss details of the three "building blocks"
(i.e.\ parton distribution function (PDF), fragmentation function (FF)
and dipole cross section), which are universal (or
process-independent) objects.  As emphasised in \cite{diquark_pict},
the scattered constituent diquarks represent a possible source of the
baryon excess measured at semi-hard $p_t$.  This forces us to
incorporate the diquark component into the first building block, i.e.\
into the PDF of a proton. We shall employ the
existing diquark PDF from Ref.~\cite{diquark_param}, which is
based on the extreme (scalar) diquark picture with $Q^2$-dependent
form-factor of the diquark, to discuss the $p_t$
spectra of hadron cross sections including scaling
violation.  Here we will introduce the breakup probability of the
diquark scattered off the CGC assuming that the cross section for
diquark scattering is damped by a $Q^2$-dependent
form-factor~\cite{diquark_pict}.  Regarding diquark fragmentation into
hadrons, phenomenological considerations using counting rules have
been applied to the data, and also more theoretical models without
scaling violations have been proposed~\cite{diquark_FF,diquark_pict}.
We will, however, not rely on these models but rather model the
diquark FFs in terms of the standard FFs with scaling violations.  As
our standard baryon FFs we employ parameterizations extracted from
JETSET simulations~\cite{XNWang}, rescaling the $Q^2$-dependence such
as to obtain a good fit to the KKP~\cite{KKP} or AKK~\cite{AKK} FFs.

Although considerable model-dependences originate from mainly the
scaling violation of baryon or diquark FFs, the inclusion of
constituent diquarks at large $x$ in the deuteron proves essential to
explain the measured baryon excess relative to standard PDF and FF
sets.  For $\pi^\pm$ and $\bar{p}$ production, on the other hand, the
diquark contribution is not so large.

This paper is organized as follows.
In the next section, we formulate the single-inclusive hadron 
production cross sections in high-energy d+Au collisions 
including mass corrections.
In Secs.~\ref{sec3} and \ref{sec4}, the predictions 
from the the model are confronted with 
experimental data for forward $\pi^\pm$ and $p/\bar{p}$ production,
respectively. A detailed discussion is given in Sec.~\ref{sec5}.
Finally, we summarize in Sec.~\ref{sec_summary}.

\section{Forward hadron production in proton-nucleus collision
within the CGC formalism}
\label{sec2}

It is well known that collinear factorization theorem is generally broken
by higher-twist (multiple scattering) effects 
in hadron-hadron collisions, although the validity
of the theorem is proved up to the twist-4 level~\cite{Collins}.
Within the CGC formalism, however,
all higher-twist effects can be included into 
one building block, i.e.\ the dipole forward scattering amplitude.
Hence, even if one considers higher-twist effects 
in p+A collisions, the predictive power of the 
theorem is still preserved~\cite{franjamal}.
It was verified also for the case with one-loop radiative corrections 
to projectile partons~\cite{aaj}.
This expression~\cite{aaj} is valid only under the condition that the target
is in the saturation or so-called ``extended geometric scaling''
regime while the projectile is dilute~\cite{ScalingViol}.
Therefore, this approach does hold
for forward rapidity kinematics which we are interested in here,
unlike the usual $k_t$-factorization approach~\cite{kt_factor}.
In this section, we shall briefly derive collinearly factorized
forms describing the single-inclusive hadron production cross section
in p+A collisions.

\begin{figure}[hbt]
\centering
\centerline{\epsfig{figure=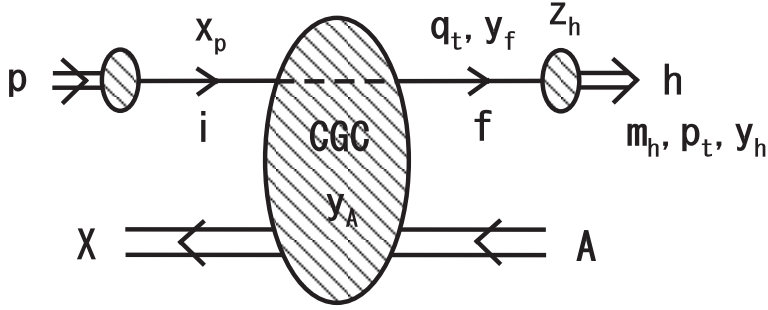,width=4in,angle=0}}
\caption{Kinematics of the $p+A \rightarrow h+X$ process.  The
incoming parton $i$ from the proton interacts with the target CGC
field of small-$x$ gluons to all orders via a $2\rightarrow 1$
process.  The outgoing parton $f$ hadronizes into a massive hadron $h$
with momentum fraction $z_h$.}
\label{fig:pA_Fdiagram}
\end{figure}

To describe the $p+A \rightarrow h+X$ process
depicted in Fig.~\ref{fig:pA_Fdiagram}, we work with light-cone momenta
in a frame where the proton has large plus component 
$P_p=(\sqrt{s/2},0,\vec{0}_t)$
and the nucleus a large minus component $P_A=(0,\sqrt{s/2},\vec{0}_t)$. 
$\sqrt{s}$ denotes the center of mass energy. 
The incoming parton $i$ carries a fraction $x_p$
of the proton momentum $P_p^+$, interacting with 
the incoming small-$x$ gluons of the
dense nucleus (the CGC) through a $2 \rightarrow 1$ process.
After the eikonal scattering process, the outgoing parton $f$
has momentum $p_f=(q_te^{y_f}/\sqrt{2},q_te^{-y_f}/\sqrt{2},\vec{q}_t)$
with rapidity $y_f$ and produces a massive hadron $h$ 
with the momentum fraction $z_h$ and mass $m_h$.
The detected hadron has a momentum 
$p_h=(m_te^{y_h}/\sqrt{2},m_te^{-y_h}/\sqrt{2},\vec{p}_t)$,
where $m_t$ is the transverse mass $m_t=\sqrt{m_h^2+p_t^2}$ and $y_h$ 
is the hadron rapidity.
The Feynman-$x$ of the produced hadron is given 
by $x_F=p_h^+/P_p^+=m_te^{y_h}/\sqrt{s}$.
In the eikonal approximation we have $z_h=p_h^+/p_f^+=x_p/x_F$ 
because of $x_p=q_te^{y_f}/\sqrt{s}$. 
The internal variables of the parton $f$, i.e. $q_t$ and $y_f$, therefore, 
can be expressed in terms of $p_t$, $m_t$, $y_h$ and $x_p$:  
$q_t=p_t/z_h=p_tx_p/x_F$ and $y_f=y_h+\log(m_t/p_t)$.
Another important internal variable is 
the momentum fraction $x_A$ carried by
the small-$x$ gluons in the nucleus. It is related to that 
of the impinging projectile parton $x_p$
in the eikonal approximation by $x_A=x_pe^{-2y_f}$ 
(see Appendix B in \cite{aaj}).
This leads to the rapidity of the gluons,
\be
y_A=\log(1/x_A)=\log(1/x_p)+2y_h+\log(m_t^2/p_t^2).
\label{eq:ya}
\ee
Here, complying with standard practice, we have defined 
the direction of the final hadron produced 
in the ``proton fragmentation region'' 
as the positive $z$-direction, i.e. $y_h > 0$.

In the framework of collinear factorization, 
the single-inclusive hadron production cross
section can be written as a convolution of 
the inclusive parton $f$ production cross section
with the PDF of the initial parton $i$, 
$f_{i/p}$, and with
the FF of $f$ into a hadron $h$, $D_{h/f}$:
\begin{eqnarray}
\lefteqn{{d\sigma(p A \rightarrow h X) \over \pi dy_h dp_t^2 d^2b}
={d\sigma(p A \rightarrow h X) \over \pi dy_h dm_t^2 d^2b}}
\nonumber\\
&=& \sum_{i,f} \int dx_p dz_h dy_f dq_t^2
\ f_{i/p}(x_p,Q_f^2)\
{d\sigma(i A \rightarrow f X) \over \pi dy_f dq_t^2 d^2b}
\ D_{h/f}(z_h,\mu_f^2)
\nonumber \\
&\times& \delta(m_t^2-M(z_h,q_t)^2)\ \delta(y_h-Y(y_f,z_h,q_t)),
\label{eq:conv1}
\end{eqnarray}
where $M(z_h,q_t)^2=m_h^2+(z_hq_t)^2$
and $Y(y_f,z_h,q_t)=y_f-\log(m_t/z_hq_t)$ 
as derived above.
$\vec{b}$ is the impact parameter with respect to the center of the nucleus. 
The factorization scale of the PDF is denoted by $Q_f$ 
and the fragmentation scale of the FF
by $\mu_f$, both of which are hereafter set to $Q_f=\mu_f=p_t$.

After integration over $y_f$, $q_t^2$ and $z_h$, 
we obtain the impact-parameter averaged 
single-inclusive hadron production cross section,
\begin{eqnarray}
{d\sigma(pA \rightarrow hX) \over \pi dy_h dp_t^2} &=& 
{1 \over 2\pi S_A} \int_{x_F}^{1} dx_p \, {x_p\over x_F} 
\int_0^{R_A} dbb 
\nonumber\\ 
&\times&
\Bigg[f_{q/p}(x_p,Q_f^2)\, N_F \left({x_p\over x_F}p_t,y_A,b\right)\,
D_{h/q}\left({x_F\over x_p}, \mu_f^2\right)
\nonumber \\
&+&
f_{g/p}(x_p,Q_f^2)\, N_A \left({x_p\over x_F}p_t,y_A,b\right)\, 
D_{h/g}\left({x_F\over x_p}, \mu_f^2\right)\Bigg].
\label{eq:conv2}
\end{eqnarray}
Here we used $d\sigma(iA \rightarrow fX)/\pi dy_f dq_t^2 d^2b
=q_f^+\delta(q_f^+-x_pP_p^+)N(q_t,y_A,b)/(2\pi)^2$, where
$N(q_t,y_A,b)$ is the scattering probability 
of dipoles from the nucleus.
$N_F$ corresponds to a projectile quark impinging 
with the target small-$x$ gluons while $N_A$ is for a projectile gluon.
The indices $q$ in $f_{g/p}$ and $D_{h/q}$ 
are summed over all quark species.
For simplicity we assume a spherical nucleus of radius $R_A$ 
with a sharp edge, where its profile function is given by
$T(b)=2\sqrt{R_A^2-b^2}$. 
$S_A$ is the transverse area of the nucleus ($=\pi R_A^2$).

The same expression as Eq.~(\ref{eq:conv2}) was derived in Ref.~\cite{aaj}
assuming massless hadrons and hence this equation 
is consistent with that of \cite{aaj} in the massless limit.
To leading $\log p_t^2$ accuracy it was verified in \cite{aaj} that 
the recoil of the projectile parton
by hard gluon radiation can be converted into the $Q^2$-evolution of
the PDF and FF according to the 
full DGLAP~\cite{dglap} evolution equations. 
The recoil effect are important 
for the interpretation of the forward-rapidity data 
from RHIC.

The dipole profiles of transverse size $r_t$
at an impact parameter $b$ are defined 
in the fundamental and adjoint representations of $SU(N_c)$, 
respectively, as
\be 
N_F(\vec{r}_t,y_A,\vec{b}) &\equiv & {1\over N_c} \, {\rm Tr_c} \,
\langle 1- V^{\dagger}(\vec{b}-\vec{r}_t/2) V(\vec{b}+\vec{r}_t/2)\rangle, 
\nonumber \\ 
N_A(\vec{r}_t,y_A,\vec{b}) &\equiv & {1\over N^2_c -1} \, {\rm Tr_c} \;
\langle 1-U^{\dagger}(\vec{b}-\vec{r}_t/2) U(\vec{b}+\vec{r}_t/2)\rangle,
\label{eq:N_FA}
\ee
where $V$ and $U$ denote Wilson lines along the light
cone~\cite{jimwlk} in the corresponding representation, and
their correlators are averaged over the color source in the nucleus.

For the actual description of the dipole profiles we rely on 
the KKT model~\cite{dima}, which is a phenomenologically 
reasonable parameterization
facilitating comparisons with experimental data,
rather than on solutions of the intricate JIMWLK equations~\cite{jimwlk},
which is not yet feasible. For a dipole in the adjoint representation,
\be
N_A(\vec{r}_t,y_A,\vec{b}) = 
1-\exp\left[-\frac{1}{4}(r_t^2 Q_s^2(y_A,b))
^{\gamma(r_t,y_A,b)}\right],
\label{NA_param}
\ee where $Q_s(y_A,b)$ is the saturation scale of the target nucleus
and $\gamma(r_t,y_A,b)$ the anomalous dimension of its gluon
distribution with saturation boundary condition.  $\gamma=1$ reduces
(\ref{NA_param}) to the Golec-Biernat-W\"usthoff saturation model
\cite{GBW} or the classical MV model \cite{MV}.  The dipole in the
fundamental representation, $N_F$, differs by a factor of $Q_s^2\to
Q_s^2\,C_F/C_A=\frac{4}{9}Q_s^2$~\cite{dima}.

The saturation scale $Q_s$ at $b=0$ is given for a nucleus of
mass number $A$ (Au(197)) as~\cite{aaj}
\be
Q_s^2(y_A,b=0) = A^{1/3} Q_0^2 \left({x_0 \over x_A}\right)^\lambda
= A^{1/3} Q_0^2 x_0^\lambda e^{\lambda y_A},
\label{Qs1}
\ee
where $Q_0\simeq 1$ GeV, $\lambda\simeq 0.3$ 
and $x_0 \simeq 3.0\times 10^{-4}$ are 
fixed by the DIS data~\cite{GBW}. 
The energy dependence of $Q_s$ is controlled through
the constant growth rate 
$\lambda=\partial \log(Q_s^2/\Lambda_{QCD}^2)
/\partial \log(1/x_A)$, which is obtained from fixed-coupling
LO BFKL evolution~\cite{IIM}.
Since the squared saturation scale has dependence on the impact parameter,
for instance, through the nuclear profile 
$T(b)=T(b=0)\sqrt{1-(b/R_A)^2}$ in the hard sphere approximation
of the nuclear target and the pointlike proton projectile,
we define the saturation scale as~\cite{Baier:2005dz}
\be
Q_s^2(y_A,b) = Q_s^2(y_A,b=0)\sqrt{1-(b/R_A)^2}.
\label{Qs2}
\ee
This naive approximation for the nuclear surface is sufficient
for minimum bias observables.

The anomalous dimension $\gamma$ is parameterized as \cite{aaj2}
\be
\gamma(r_t,y_A,b) &=& \gamma_s + 
(1-\gamma_s)\,
\frac{|\log(1/r_t^2Q_s^2(y_A,b))|}
{\lambda y_A+|\log(1/r_t^2Q_s^2(y_A,b))|+d\sqrt{y_A}},
\label{eq:gam_new}
\ee where $\gamma_s\simeq 0.627$ is the anomalous dimension for BFKL
evolution~\cite{bfkl} with saturation boundary condition, i.e.\ for
evolution along the saturation line \cite{IIM}, and $d$ is a free
parameter which is fitted to experimental data.  
Throughout this paper, we will make replacement
$\gamma(r_t,y_A,b)\rightarrow \gamma(1/q_t,y_A,b)$.
As indicated in
\cite{aaj2}, this parameterization of $\gamma$ stays within
$\gamma=0.6\sim 0.8$ at large rapidity $y_h$ over a comparatively wide
range of $p_t=1\sim 5$ GeV~\footnote{As emphasized in \cite{aaj2},
this remains inside the (at least, extended) geometric scaling
regime. When going beyond the scaling regime like higher-$p_t$ or
lower rapidities, Eq.~(\ref{eq:gam_new}) assumes that crossover from
the scaling regime to the perturbative one is very slow and smooth. A
similar behavior of $\gamma$ was discussed in Ref.~\cite{BKW}}.

Below we focus on the minimum-bias cross section obtained by
impact-parameter averaging of Eq.~(\ref{eq:conv2}). 
Since in the integrand of (\ref{eq:conv2}) the impact-parameter
dependence is carried only by the saturation scale,
we take the average value of $Q_s^2(b)$ with respect to $b$
instead of integrating (\ref{eq:conv2}) over $b$:
\be
\langle Q_s^2(b)\rangle\equiv {\pi \over S_A}\,
\int_0^{R_A^2} db^2 Q_s^2(y_A,b)
= {2 \over 3}\,Q_s^2(y_A,b=0).
\label{eq:Qs_ave}
\ee 
We have used Eq.~(\ref{Qs2}) where this approximation is valid to
good accuracy \cite{Baier:2005dz}~\footnote{Refs.~\cite{aaj,aaj2} 
employ an effective saturation scale for minimum bias collisions such
as $\langle Q_s^2(b)\rangle=A_{eff}^{1/3} Q_0^2 (x_0/x_A)^\lambda$
with $A_{eff}=18.5$, where the factor $A_{eff}^{1/3}$ is by $\sim 30\%$
smaller than the corresponding factor $(2/3)A^{1/3}$ in (\ref{eq:Qs_ave}).}.

Then, Eq.~(\ref{eq:conv2}) reads
\begin{eqnarray}
{d\sigma_{m.b.}(pA \rightarrow hX) \over \pi dy_h dp_t^2} 
&=& 
{1 \over (2\pi)^2} \int_{x_F}^{1} dx_p \, {x_p\over x_F}
\Bigg[f_{q/p}(x_p,Q_f^2)
N_F\left({x_p\over x_F}p_t,y_A,\langle Q_s^2(b)\rangle\right)
D_{h/q}\left({x_F\over x_p}, \mu_f^2\right)
\nonumber \\
&+&
f_{g/p}(x_p,Q_f^2) 
N_A\left({x_p\over x_F}p_t,y_A,\langle Q_s^2(b)\rangle\right)
D_{h/g}\left({x_F\over x_p}, \mu_f^2\right)\Bigg],
\label{eq:conv3}
\end{eqnarray}
where the Fourier transform of the dipole profile functions is given by
\be
N_{A,F}(q_t,y_A,\langle Q_s^2(b)\rangle) 
&=& - \int d^2r_t \;e^{i\vec{q}_t\cdot\vec{r}_t}
N_{A,F}(r_t,y_A,\langle Q_s^2(b)\rangle)
\nonumber\\
&=& - 2\pi\int_0^\infty dr_t \;r_t \;J_0(r_t \,q_t)\;
N_{A,F}(r_t,y_A,\langle Q_s^2(b)\rangle).  
\label{FTdipole}
\ee

\section{Forward $\pi^\pm$ production in d+Au collision at RHIC}
\label{sec3}

We are now in the position of applying our results (\ref{eq:conv3})
to deuteron-gold collisions at RHIC energy $\sqrt{s}=200$ GeV, 
at large rapidity ($y_h=3.0$). We first address
minimum bias $\pi^\pm$ production observed 
by BRAHMS collaboration~\cite{Debbe}.
This process is dominated by valence quarks since $x_p \gsim 0.1$ on the
deuteron side and hence might be affected
by initial quark correlation (like diquarks) with large $x$ momentum 
inside deuteron.

In this section, our interest is twofold:
First, we check the validity of our leading order 
CGC formalism using standard 
parameterizations of the PDF and the FF, and determine the free parameter
$d$ as well as the $K$-factor prescribed to NLO corrections to reproduce well
the experimental data.
Second, using those very same parameters ($d$ and $K$) we investigate
the diquark contribution.

\begin{figure}[hbt]
\centering
\centerline{\epsfig{figure=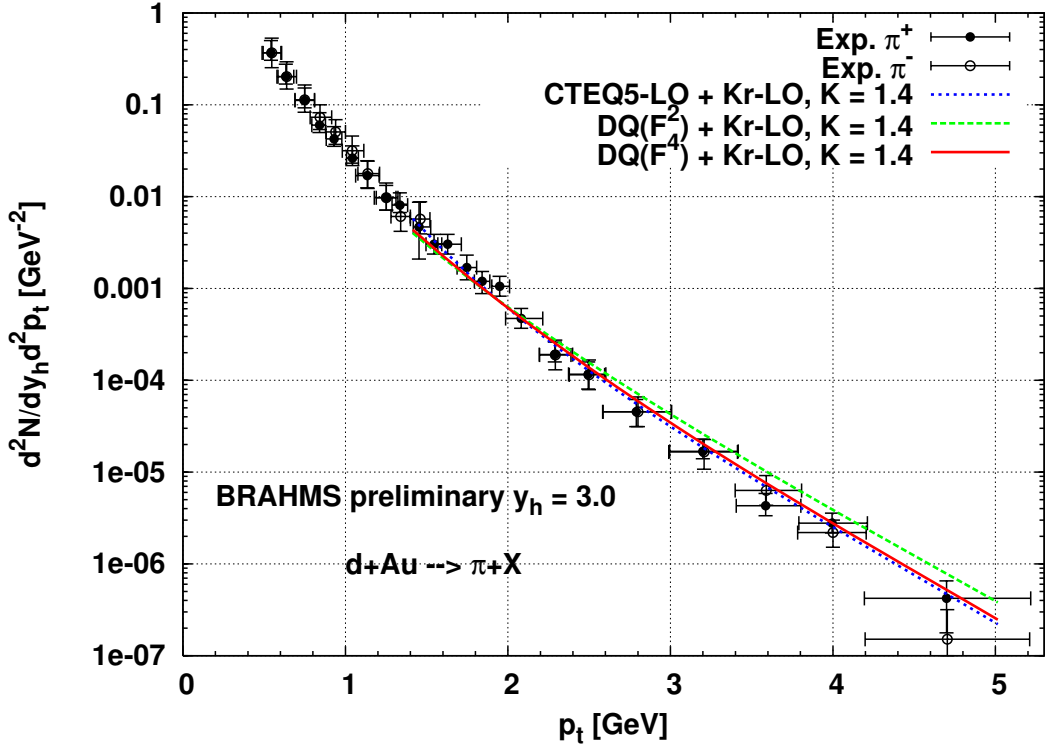,width=4in,angle=-90}}
\caption{$p_t$ spectra of $\pi^+$ and $\pi^-$ 
from d+Au collisions compared to
BRAHMS preliminary minimum bias data ($y_h=3.0$).
The lines show the CGC results with the LO Kretzer FFs and $K=1.4$, 
where either the LO CTEQ5 or diquark PDFs are used.}
\label{fig:dAuf30_chargedpi}
\end{figure}

For a deuteron projectile, we treat its parton distributions
as a simple superposition of those in a proton and a neutron 
without any nuclear modification, which
indeed has negligible effects (less than $5\%$) for the deuteron.
The parton distributions in the neutron are obtained through
isospin symmetry, $f_{u,\bar{u}/p}=f_{d,\bar{d}/n}$
and $f_{d,\bar{d}/p}=f_{u,\bar{u}/n}$, and other parton species
have the same distribution as those in the proton~\cite{wv}.
This isospin symmetry may affect the final pion productions
composed of flavor $SU(2)$ light quarks,
where it leads to no difference between 
light $\pi^+(u\bar{d})$ and $\pi^-(\bar{u}d)$ productions to good accuracy.
Actually, such behavior can be seen for $p_t \lsim 4$ GeV
in the BRAHMS data plotted in Fig.~\ref{fig:dAuf30_chargedpi}.

We first use the leading order (LO) CTEQ5 PDFs~\cite{cteq}.  Then, we
have several choices for the FFs, e.g. the Kretzer~\cite{Kretzer},
KKP~\cite{KKP} and AKK~\cite{AKK} sets. The FFs by Kretzer assume that
the charge-conjugation symmetry $D_{q/\pi^\pm}=D_{\bar{q}/\pi^\mp}$
holds at low input scale.  Both KKP and AKK FFs (the latter provide
flavor-dependent FFs of light quarks as an update of KKP) provide only
the average of charged pion FFs.  Therefore, all three FFs give
the same neutral pion FF and should show almost the same
$p_t$-spectra at LO.

We checked that the result with the LO Kretzer FFs gives a very similar
curve to that of the LO KKP FFs (but different $K$-factor)
over the range $1\lsim p_t \lsim 5$ GeV, with
a pion mass of $m_\pi=0.14$ GeV.
The result with the LO Kretzer FFs is in good agreement with the data 
at $d=0.6$ and $K=1.4$, while the LO KKP FFs require $d=0.6$ and $K=1.0$.
The main difference is due to the way that gluons fragment to pions;
its contribution in the KKP set is larger than that 
in the Kretzer one~\cite{STAR_Adams}.
The sensitivity of the results to the value $d$ is not so large 
in the range of $d=0.6\sim1.2$ and is visible only at high $p_t$, 
where the data however have large errors.
In Fig.~\ref{fig:dAuf30_chargedpi} 
we show the LO Kretzer result with $K=1.4$ as the dotted line.

Next, in order to investigate the diquark contribution to this process,
we make use of the PDFs given by the Stockholm diquark
model~\cite{diquark_param}. It assumes that only scalar diquarks 
are genuine bound states and other axial-vector diquarks are negligible.
As is well known, the diquark is classified into $\bar{3}_c$ and
$6_c$ channels in color $SU(3)$. 
The scalar diquark belongs to the $\bar{3}_c$
channel while the axial-vector diquark is in the $6_c$ representation.
The reason why we adopt only the former is that the one-gluon
exchange between quarks is attractive in the $\bar{3}_c$ channel
while repulsive in the $6_c$ channel. Thus, for a first estimate
it is rather plausible to focus on the more tightly bound scalar-diquark.
This parameterization is obtained from a fit to the proton structure function
$F_2^p$ observed in high energy e+p collisions at SLAC, BCDMS and EMC,
over the wide range $1 < Q^2 < 200$ GeV$^2$.

Ref.~\cite{diquark_param} has two possible parameterizations 
which differ by the choice of a mass scale in the diquark form-factor.
We choose the parameter set with $M^2=10$ GeV$^2$ and a dipole
form-factor
\be
F^2(Q_f^2)={1 \over (1+Q_f^2/M^2)^2}.
\label{form-factor}
\ee 
It ensures that with larger $Q_f^2$ all diquarks are resolved and
the usual quark PDFs are recovered. 
The scale $M$ is related to the internal binding energy of the
diquark and $M^2=10$ GeV$^2$ corresponds to a diquark radius 
of about 0.2 fm, which is obtained from the mean-square radius
$\langle r^2\rangle=6dF^2(Q_f^2)/dQ_f^2|_{Q_f^2=0}$. 
It yields larger contribution from diquark
component due to its tightly bound state than that of $M^2=3$ GeV$^2$.
The parameterization for each parton, then, 
is given as~\cite{diquark_param}
\be
x_p u(x_p,Q_f^2) &=& \frac{(0.52+0.37s)x_p^{0.45+0.12s}
(1-x_p)^{2.2+2.39s}-x_p^{0.71-0.02s}(1-x_p)^{7.3-0.51s}}
{(0.52+0.37s)B(0.45+0.12s,3.2+2.39s)-B(0.71-0.02s,8.3-0.51s)}
\nonumber\\
&+& \frac{x_p^{0.9-0.83s}(1-x_p)^{5.0-1.88s}}{B(0.9-0.83s,6.0-1.88s)}
\,[1-F^2(Q_f^2)],
\nonumber\\
x_p d(x_p,Q_f^2) &=& \frac{x_p^{0.9-0.83s}(1-x_p)^{5.0-1.88s}}
{B(0.9-0.83s,6.0-1.88s)}\,[1-F^2(Q_f^2)],
\nonumber\\
x_p s(x_p,Q_f^2) &=& x_p \bar{u}(x_p,Q_f^2) = x_p \bar{d}(x_p,Q_f^2)
= x_p \bar{s}(x_p,Q_f^2)
\nonumber\\
&=& (0.35-0.06s)(1-x_p)^{6.89+0.75s},
\nonumber\\
x_p f_{DQ}(x_p,Q_f^2) &=& \frac{x_p^{0.93-0.52s}(1-x_p)^{1.5-1.1s}}
{B(0.93-0.52s,2.5-1.1s)}\,F^2(Q_f^2),
\label{diquark}
\ee where we assumed flavor $SU(3)$ symmetry for the sea quark
distributions, and $f_{DQ}$ denotes the diquark distribution,
$B(\mu,\nu)$ Euler's Beta function
and 
$s=\log[\log(Q_f^2/\Lambda^2)/\log(Q_0^2/\Lambda^2)]$
with $\Lambda=0.2$ GeV and $Q_0^2=4$ GeV$^2$.

When one views the d+Au collisions in this diquark picture, one faces
other theoretical uncertainties besides modelling of the diquark
distribution in the deuteron: the breakup of diquarks scattered off
the CGC and the diquark fragmentation into hadrons.  The former
depends on the collision dynamics between diquarks with a finite size
and the CGC.  To take into account this breakup probability in a
simple way, we express it as the dipole form-factor of
(\ref{form-factor}) with $M^2=10$ GeV$^2$ as done in
Ref.~\cite{diquark_pict}.  This probability goes to zero in the
high-$p_t$ limit ($Q_f=p_t$), 
and then the diquark bound state is completely
broken inside the CGC, where the constituent $u$ and $d$ quarks fragment
independently.  

Although so far various models for the diquark fragmentations have
been proposed~\cite{diquark_FF}, they have focused mainly on how the
FFs behave as a function of $x$, leaving aside scaling violation by
the $Q^2$ evolution.  For our present investigation of $p_t$-spectra
of pions and baryons from diquarks it is important to account for the
scaling violation.  For this purpose, as a naive model estimate we
express the diquark FFs in terms of the LO Kretzer FFs, which describe
fragmentation into charged hadrons separately~\footnote{The standard
FFs like the Kretzer set describes fragmentations from a color triplet
parton.  Since the scalar diquark is in the color
anti-triplet representation, it may hadronize similarly to a parton, at least
as far as color is concerned, if the size of diquark is ignored.}.  We
approximate the $(ud)$-diquark FFs into charged pions as \be
D_{\pi^+(u\bar{d})/DQ(ud)}(z_h,\mu_f^2) &=&
D_{\pi^+/d}^{Kr-LO}(z_h,\mu_f^2)F^2(\mu_f^2) \nonumber\\ &+&
[D_{\pi^+/u}^{Kr-LO}(z_h,\mu_f^2)+D_{\pi^+/d}^{Kr-LO}(z_h,\mu_f^2)]
(1-F^2(\mu_f^2)), \nonumber\\ D_{\pi^-(\bar{u}d)/DQ(ud)}(z_h,\mu_f^2)
&=& D_{\pi^-/u}^{Kr-LO}(z_h,\mu_f^2)F^2(\mu_f^2) \nonumber\\ &+&
[D_{\pi^-/d}^{Kr-LO}(z_h,\mu_f^2)+D_{\pi^-/u}^{Kr-LO}(z_h,\mu_f^2)]
(1-F^2(\mu_f^2)),
\label{dq_pion}
\ee where we multiplied the FFs by the breakup probability.  The first
term is the contribution where the valence quarks of the produced
pions do not include neither one of the quarks from the scattered
diquark, and thus the pion production occurs via quark-antiquark pair
creation from the vacuum, like $D_{\pi^+/d}$ or $D_{\pi^-/u}$.  The
second contribution corresponds to diquark breakup followed by
independent fragmentation of each quark into pions.

In Fig.~\ref{fig:dAuf30_chargedpi} we plot the diquark result, using
Kretzer FFs both for the diquarks as well as for the single partons,
as the dashed line (with $d=0.6$ and $K=1.4$).  This curve, however,
has a problem, since it should approach the result with CTEQ5+Kretzer
(dashed line) at high $p_t$ due to $F^2(p_t^2) \rightarrow 0$; that
is, the diquark contribution should vanish in this limit and the
single parton picture should be restored again, but apparently the
former is much harder than the latter.  To improve this unfavorable
behavior, we simply change the dipole form-factor $F^2(p_t^2)$ to
$F^4(p_t^2)$ for both the PDFs and the breakup probabilities.
This result is shown as the solid line, which is softer at high $p_t$
and is in good agreement with the CTEQ5+Kretzer result, especially at
high $p_t$.  Since this change of the form-factor increases the
diquark radius from 0.2~fm to 0.3~fm~\footnote{This radius is still smaller
than the other parameter set of \cite{diquark_pict} with $M^2=3$ GeV$^2$,
where $\sqrt{\langle r^2\rangle}\simeq 0.4$ fm.}, the new form-factor
becomes more sensitive to $p_t$ and the diquarks are broken up more
easily.  The solid line is still somewhat harder than the dotted one
at low $p_t$, but it remains a good fit to the data.  Thus, the
result in the diquark picture can reproduce well the data within
error bars.  As expected, this implies that the diquark contribution to
pion production is not very large.

\section{Forward $p$ and $\bar{p}$ production}
\label{sec4}

\begin{figure}[hbt]
\centering
\centerline{\epsfig{figure=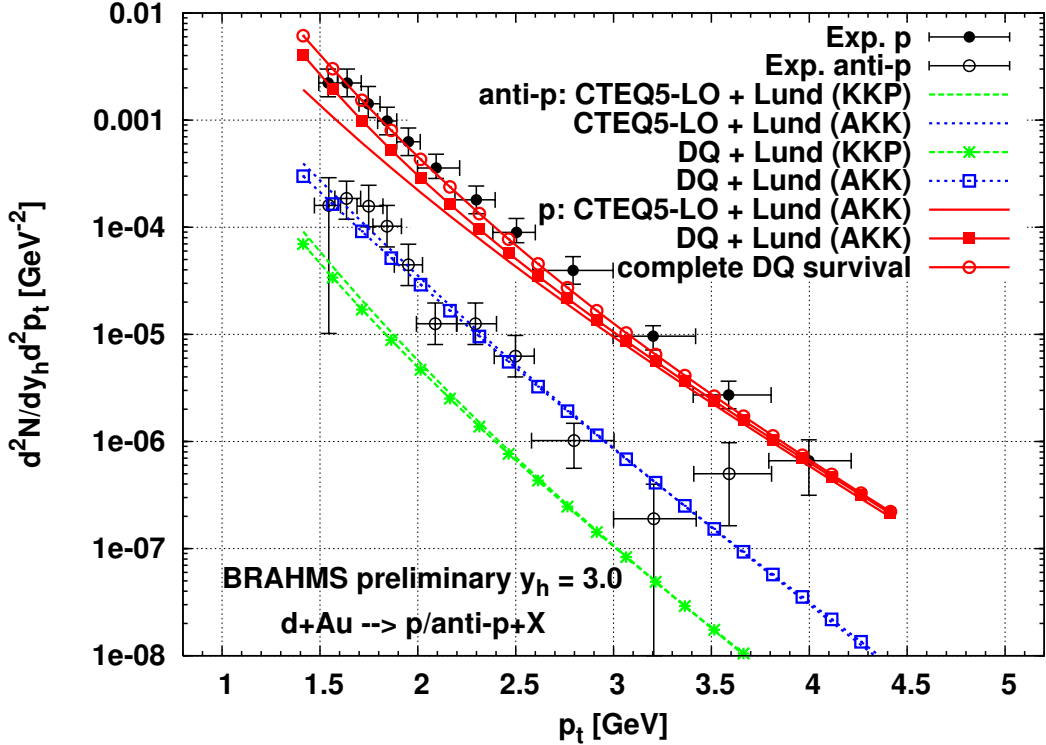,width=4in,angle=-90}}
\caption{$p_t$ spectra of $p$ and $\bar{p}$ in d+Au collisions
from BRAHMS preliminary minimum bias data ($y_h=3.0$) compared to
computations within the CGC formalism ($K=1.0$).
The lines show the results with the common LO CTEQ5 PDFs 
and the Lund FFs rescaled to either the KKP or AKK FFs with $K=1.0$. 
The lines with symbols show the results with the diquark PDFs.}
\label{fig:dAuf30_ppbar}
\end{figure}

In this section we examine forward proton or anti-proton production
in d+Au collisions at RHIC and apply Eq.~(\ref{eq:conv3}) to
obtain the cross sections within the same setup 
as for pion production (discussed above).
At the RHIC energy $\sqrt{s}=200$ GeV and
in the deuteron fragmentation region ($y_h=3.0$), 
we first show the BRAHMS preliminary data~\cite{Debbe} for 
minimum bias cross sections of $p$ (solid circle) 
and $\bar{p}$ (open circle) production in Fig.~\ref{fig:dAuf30_ppbar}.

Unlike for pions, charge symmetry is broken more strongly for $p$,
$\bar{p}$ production. The proton cross section exceeds that of
anti-protons by one order of magnitude for $1.5\lsim p_t \lsim 3.5$
GeV.  This feature can also be seen in p+p collisions at large
rapidity and high $p_t$~\cite{Debbe}.  Such a large asymmetry between
them, and in particular the high abundance of protons, may be
associated with the production mechanism involving the fragmentation
of diquarks into protons.

To our knowledge, the best parameterization of baryon FFs handling the
$Q^2$ evolution to some extent is given in Ref.~\cite{XNWang}, where
the FFs are obtained by parameterizing results computed by Monte Carlo
simulations in the Lund string model, evaluated at invariant mass
$W=2Q_f$.  The advantage of using this parameterizations is that we can
treat proton and anti-proton production separately while the KKP and
AKK FFs only provide averages. Moreover, the Lund string model
JETSET also models mass effects {\em in the fragmentation} which are
important when the momentum of the outgoing parton is not much larger
than the mass of its daughter hadrons.  However, the $Q_f^2$
dependence of these Lund FFs $D_{B/q}(z_h,Q_f^2)$ has not been
implemented very carefully: Ref.~\cite{XNWang} only requested that the
average multiplicity $\int_0^1dz_h D_{B/q}(z_h,Q_f^2)$ satisfies the
$Q_f^2$ dependence of the experimental data.  This treatment would not
describe correctly scaling violations, especially in the high-$z_h$ regions,
where the FFs change steeply.

\begin{figure}[hbt]
\centering
\centerline{\epsfig{figure=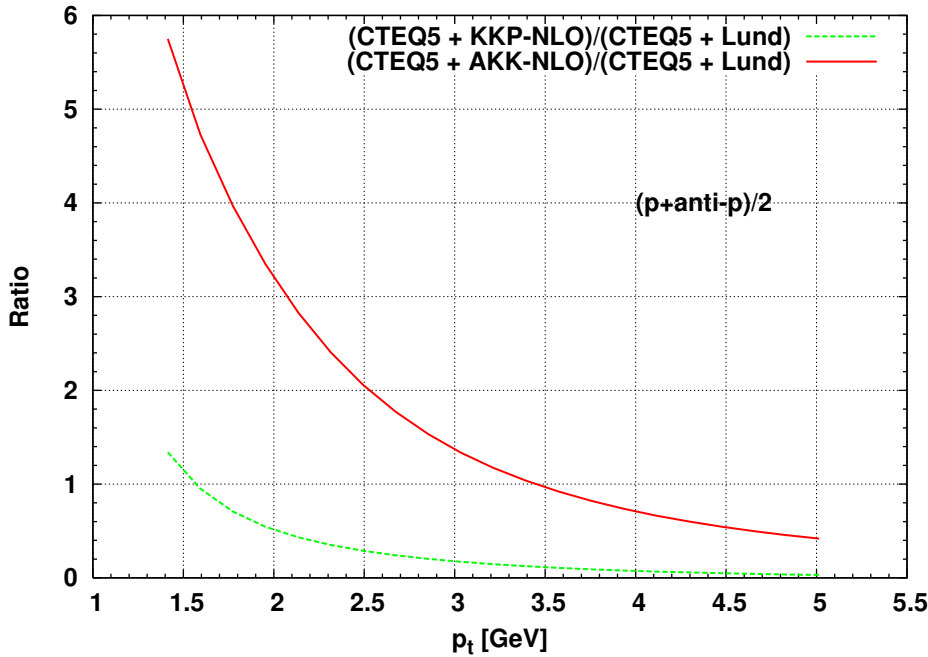,width=3.5in,angle=-90}}
\caption{Summed $p+\bar{p}$ yields obtained with the NLO KKP or NLO
AKK FFs divided by the Lund FFs.}
\label{fig:dAuf30_ppbar_ave}
\end{figure}

To check the $Q_f^2$ behavior, we computed the average
$p_t$-distribution of protons and anti-protons, $(p+\bar{p})/2$, and
compared with the results obtained with the NLO KKP and NLO AKK FFs,
using the LO CTEQ5 PDF for all three FF sets~\footnote{ The reason why
we employ the NLO KKP or NLO AKK FFs is as follows: Recent STAR data
show that high-$p_t$ $p+\bar{p}$ yields around midrapidity in d+Au
collisions at $\sqrt{s}=200$ GeV agree better with NLO pQCD
calculations using the AKK rather than the KKP FFs~\cite{STAR_Adams}.
The latter is known to underestimate the data by one order of
magnitude and so it would be best to adopt the AKK FFs for baryon
production.  Unfortunately, the AKK FFs only provide a NLO
parameterization.  The use is inconsistent within our LO formalism, but
despite such inconsistency, the use of AKK FFs for baryon productions
is of importance to reproduce the yields. We also checked that the NLO
KKP FFs are quite similar to the LO KKP FFs, but the former is a bit
softer than the latter at high $p_t$.}.
The ratios
KKP-NLO/Lund and AKK-NLO/Lund are 
shown in Fig.~\ref{fig:dAuf30_ppbar_ave},
where the baryon mass is set to $m_{p,\bar{p}}=0.938$ GeV and 
we choose $d=0.6$ and $K=1.0$.

We see that the ratios change steeply as a function of $Q_f=p_t$
and that the $Q_f^2$ dependence of the Lund FFs is
significantly harder than those of KKP and AKK sets.
In more detail, the cross section in the Lund model is 
larger than that with the KKP set
for $p_t\gsim 1.5$ GeV, while the behavior is opposite for the AKK set
at $p_t\lsim 3.5$ GeV.
This forces us to improve the Lund model FFs
to enforce a $Q_f^2$-dependence that matches that of the KKP or AKK FFs.
In this figure we can also see that the KKP FFs are smaller
by a factor of $5\sim 12$ than the AKK FFs 
in the region $1.5 \lsim p_t \lsim 5$ GeV 
and the deviation becomes larger at higher $p_t$.

For that purpose, we fit the ratios using each function via
\be
{\rm KKP(NLO)/Lund} &=& {1\over [-0.01+(p_t/1.56)^{1.36}]^{2.01}},
\nonumber\\
{\rm AKK(NLO)/Lund} &=& {1\over [0.12+(p_t/3.70)^{1.91}]^{1.38}}.
\label{fit}
\ee 
These functions provide rather reasonable fits.
Multiplying the Lund FFs by the fitting functions~(\ref{fit}), 
we then rescale the $Q_f^2$ dependence of the FFs.
We shall employ the same rescaling functions 
for both $p$ and $\bar{p}$ FFs.

For the diquark fragmentation into baryons we construct 
the following model in the same way as for pion production:
\be
D_{p(uud)/DQ(ud)}(z_h,\mu_f^2) &=& 
D_{\pi^+(u\bar{d})/\bar{d}}^{Kr-LO}(z_h,\mu_f^2) F^2(\mu_f^2)
\nonumber\\
&+& [D_{p/u}^{Lund}(z_h,\mu_f^2) + D_{p/d}^{Lund}(z_h,\mu_f^2)]
(1-F^2(\mu_f^2)),
\nonumber\\
D_{\bar{p}(\bar{u}\bar{u}\bar{d})/DQ(ud)}(z_h,\mu_f^2)
&=& {1 \over 2}[D_{\bar{p}/u}^{Lund} + D_{\bar{p}/d}^{Lund}]F^2(\mu_f^2)
\nonumber\\
&+& [D_{\bar{p}/u}^{Lund}(z_h,\mu_f^2) + D_{\bar{p}/d}^{Lund}(z_h,\mu_f^2)]
(1-F^2(\mu_f^2)),
\label{dq_baryon}
\ee where $D^{Kr-LO}$ and $D^{Lund}$ denote the LO Kretzer and Lund sets
respectively.  The first terms express the diquark fragmentation as a
single entity and the second ones do as a system consisting of
independent valence quarks.  The first term of the proton FF is based
on the conjecture that the fragmentation of the $(ud)$-diquark
proceeds via pick-up of a $u$-quark from the vacuum, similar to that
of $\bar{d}$ into $\pi^+$, if the finite size of diquark is neglected.
The first term of the anti-proton production treats a diquark as a
single entity (like $u$ or $d$ quarks) and hence takes their average.

In Fig.~\ref{fig:dAuf30_ppbar}, we plot the results of Eq.~(\ref{eq:conv3})
rescaled to the KKP or AKK FFs, and using the form-factor $F^4$.
For anti-protons,
we show a total of four curves with the CTEQ5
or diquark PDFs,
rescaled to either the KKP (dashed) or AKK (dotted) FFs.
As indicated above, the KKP set significantly underestimates
the anti-proton data by about a factor 10,
independent of the choice of PDFs. The AKK FFs fit
better although the data appears somewhat softer than our results.
The difference between the CTEQ5 and diquark PDFs is generically small but
increases at low $p_t$, similarly to
the pion case discussed in Sec.~\ref{sec3}. 
Both these curves are in good agreement with the data 
within the large error-bars.

For protons, we first show the CTEQ5 result rescaled to
the AKK set as the solid line, which underestimates the data
by a factor $2\sim 3$ in the low-$p_t$ region, 
but approaches the data at high $p_t$.
Next, using the same FFs, the diquark result is plotted as 
the square+solid line, which enhances the cross section 
at low $p_t$, but still underestimates the data.
This feature is, however, unlike the cases of anti-protons or pions and 
the enhancement of the cross section at low $p_t$ is favored by the data.
As a somewhat extreme case we also plot a curve corresponding to
``complete diquark survival'' by setting 
the breakup probability to zero (i.e.\ $F^4(\mu_f^2)=1$)
as the open-circle+solid line. 
This result is very close to the data in the low-$p_t$ region.

\begin{figure}[hbt]
\centering
\centerline{\epsfig{figure=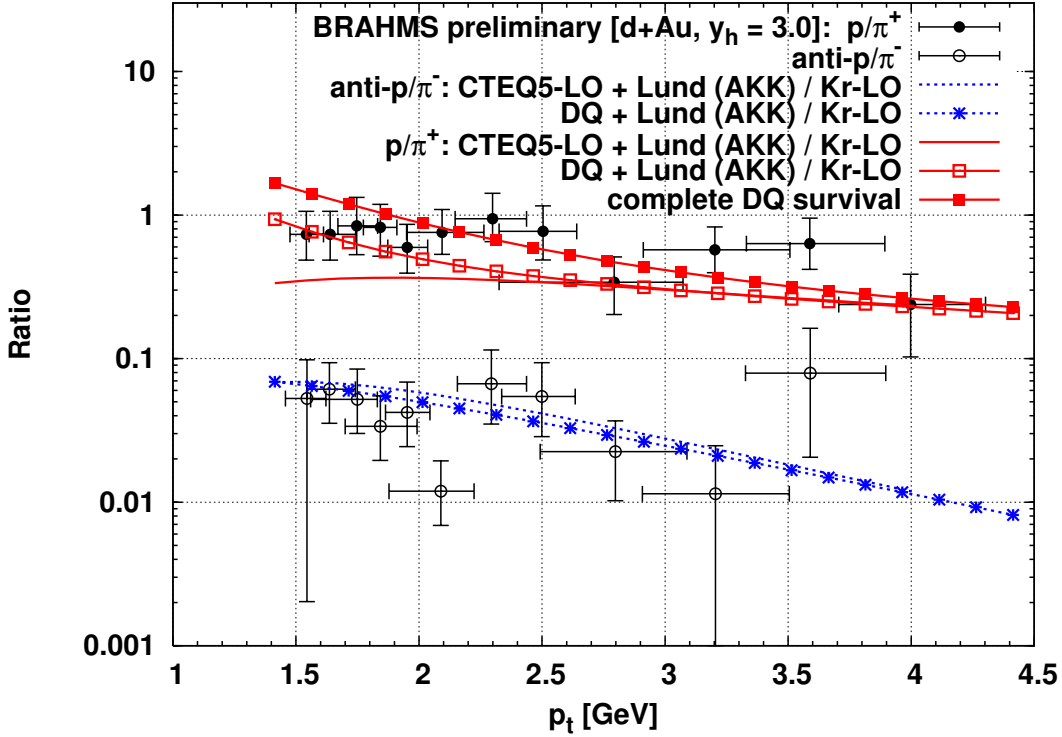,width=4in,angle=-90}}
\caption{Ratios of $p/\pi^+$ and $\bar{p}/\pi^-$. 
The experimental points were obtained
by taking the ratios of the corresponding data 
from Figs.~\ref{fig:dAuf30_chargedpi} and \ref{fig:dAuf30_ppbar}.
The lines show the results with the common LO CTEQ5 PDFs 
for both baryons and pions, where the former uses the Lund FFs
rescaled to the AKK FFs with $K=1.0$ 
and the latter the LO Kretzer FFs with $K=1.4$. 
The lines with symbols show the results with the diquark PDFs.}
\label{fig:dAuf30_proton_ratio}
\end{figure}
The $p/\pi^+$ and $\bar{p}/\pi^-$ ratios will be relevant for
investigating the $p_t$ dependence of baryon production more
precisely.  In Fig.~\ref{fig:dAuf30_proton_ratio} we plot the data
extracted by taking the ratios of the corresponding data points from
Figs.~\ref{fig:dAuf30_chargedpi}, \ref{fig:dAuf30_ppbar}.
Here, as seen in Fig.~\ref{fig:dAuf30_chargedpi} 
there are no data points of the $\pi^-$ in the range $1.5\lsim
p_t \lsim 2.1$ GeV. To interpolate the lack in this interval (6 points), 
we substitute the corresponding data points of the $\pi^+$.  
This is reasonable because for a deuteron projectile isospin symmetry seems
to work well at low $p_t$ as discussed above.  This figure shows that
the $p/\pi^+$ ratio is almost $0.6 \sim 0.9$ which is almost flat, 
while the $\bar{p}/\pi^-$ ratio is $0.04 \sim 0.07$, 
smaller by one order of magnitude, where at $p_t\lsim 2.5$ GeV we see no
significant $p_t$-dependence, but above this point
the $p_t$-distribution starts to decrease largely with $p_t$.

For the $\bar{p}/\pi^-$ ratio, we plot two results: The dotted line
uses the common LO CTEQ5 PDFs for both $\bar{p}$ and $\pi^-$, and the
Lund FFs rescaled to the AKK set for $\bar{p}$ and the Kretzer FFs for
$\pi^-$, with different $K$-factors: $K=1.0$ for $\bar{p}$ and $K=1.4$
for $\pi^-$.  The cross-dotted line replaces the PDFs by the diquark
set, with otherwise the same ``ingredients''.  These curves explain
the data quantitatively, especially at low $p_t$, with small
discrepancies between the CTEQ5 and diquark PDFs. The monotonic decrease
with $p_t$ is supported by the data except the last data point.

For the $p/\pi^+$ ratio, we show the same curves~\footnote{This ratio
is above unity around $p_t=1.5$ GeV, because of the relatively
large underestimate of our $\pi^+$ cross section as compared to the $\pi^+$
data.}.  First, we employ CTEQ5 PDFs with the Lund FFs rescaled to
the AKK set for $p$ and to the Kretzer FFs for $\pi^+$ (solid line).
Second, we switch to the diquark PDFs with the same sets of FFs.  The
experimental data cannot be explained by the use of the CTEQ5 PDFs any
more, which is too low by a factor of $2 \sim 3$ at low $p_t$, as
already seen in Fig.~\ref{fig:dAuf30_ppbar}.  The result is almost
flat over a wide range of $p_t$ and seems to work well only in the
high-$p_t$ region.  Switching to the diquark PDFs may explain at least
partly the large baryon excess seen in the data at low $p_t$. Diquark
effects disappear quite rapidly with increasing $p_t$ which is not
really supported by the data. The prediction of the ``surviving
diquark'' picture is moderately larger at low $p_t$ and perhaps offers
a reasonable explanation of the data.

\section{Discussion}
\label{sec5}

The physical process of d+Au collisions illustrated in
Fig.~\ref{fig:pA_Fdiagram} shows that the cross section is factorized
into three parts, i.e.\ into a convolution of the form
$f_{q,g/d}\otimes N_{F,A} \otimes D_{h/q,g}$ like in
Eq.~(\ref{eq:conv3}) (refer to ref.~\cite{aaj} for verifying this
factorization to leading logarithmic accuracy~\footnote{The process
violating the factorization like the second rescattering of the hard
parton with the target after the first rescattering and the succeeding
gluon radiation vanishes in the high energy limit~\cite{fgjjm,jjmyk},
at least if we calculate in light-cone gauge.}).  These three
functions are universal, independent of the processes and associated
with infrared hadron or nuclear structures, for example with respect
to the fact that we employ the same forms of $N_{F,A}$ for meson
($D_{M/q,g}$) and baryon ($D_{B/q,g}$) production.  In other words,
the $p_t$ spectra may be determined by universal dipole profiles,
irrespective of detected hadron species.
These dipole profiles $N_{F,A}$ describe the interaction of hard
partons scattering off the fields of nucleus. Bremsstrahlung occurring
either ``before'' or ``after'' the propagation through the nucleus
yields logarithmic radiative corrections, which are absorbed into the
DGLAP $Q^2$-evolution of $f_{q,g/d}$ and $D_{h/q,g}$ \cite{aaj}.

As seen from Figs.~\ref{fig:dAuf30_chargedpi}, \ref{fig:dAuf30_ppbar},
the data displays a common power-law behavior over the range of
$1\lsim p_t \lsim 5$ GeV, not an exponential form predicted by the
parton recombination model~\cite{rudi}.  In more detail, it is close
to the exponential until $p_t\sim 2-3$~GeV, but above this it appears
rather close to a power-law.  In the CGC formalism, the dipole profile
displays a power-law behavior for the $p_t$-distribution at high
rapidity, whose specific shape is governed by the anomalous dimension
$\gamma$~\cite{aaj,aaj2}.

In fact, our dipole profiles give even slightly harder $p_t$-distributions
than the data for both $p$ and $\bar{p}$ if we employ the CTEQ PDFs.
If the factorization and the CGC framework work well, however, 
such a deviation from the data will be due mainly to
the $Q^2$-evolution FFs (here the Lund baryon FFs rescaled by the AKK FFs).
If this is the case, we should determine more carefully 
the scaling violation of the baryon FFs in the Lund string scheme.
Work on new parameterizations by means of Lund JETSET simulation 
will be reported elsewhere~\cite{dhjn}.

For $p$ production, a remarkable point is that the diquark picture
does have a noticeable effect on the spectra at large rapidity. In
fact, for the specific parameterizaiton employed here, the slope is
even steeper than seen in the data.  This comes from mainly the
$p_t$-dependence in the form-factors, which are used both in the PDFs
and in the breakup probability of diquarks.  As a reference, we
considered the case of completely pointlike diquarks, which are not
broken by scattering off the CGC at all.  At a first glance this
result is closer to the data compared to the results with breakup.
However, this should be taken with caution, because this "surviving
diquark'' picture maximizes the contribution from diquarks and is
inconsistent with the baryon suppression scenario in the deep
saturation regime~\cite{DGS}.

For a more quantitative comparison to experimental data we require
more precise information of the $Q^2$-dependence of diquark
distribution and fragmentation functions (which can not be measured in
$e^+e^-$ annihilation like ordinary quark FFs).  It might also be possible that
the deuteron is composed of three-diquark system~\cite{tri_diquark}.
In the present formulation the third constituent diquark has not been
considered. This contribution may lead to additional baryon formation.

\section{Summary} 
\label{sec_summary}

We have focused on forward $p$ and $\bar{p}$ production in d+Au
collisions using universal dipole profiles developed in
\cite{aaj,aaj2} within the CGC formalism.  This formalism is in a good
agreement with the high $p_t$-spectrum of forward pions at $y=3.0$,
where its power-law behavior is described well by the dipole profile
in the formalism.  However, the abundance of protons can not be
explained by the usual PDFs (like CTEQ5) and Lund FFs (rescaled by AKK
FFs), while that of anti-protons is consistent with our results.
Similarly, large deviations of the $p/\pi^+$ ratio from Pythia
simulations is already observed for forward p+p
collisions~\cite{Debbe}.  One possible explanation for this extra
production of protons at $p_t=1\sim 3$ GeV compared to standard
PDFs is the direct formation of protons from (dominantly scalar)
diquarks, whose contribution is incorporated into our
formulation, including a diquark form-factor and fragmentation into
hadrons.  This diquark contribution is essential only for proton
production but is less important for pions and anti-protons.

\section*{Acknowledgements}
We would like to give A.~Dumitru special acknowledgement for useful comments 
and his careful reading of the manuscript.
We would also thank Y.~Nara for fruitful discussions 
in the whole process of accomplishing this work, J.~Jalilian-Marian
for encouraging comments, and H.-J.~Drescher for
providing the faster computer program for computing the 
Fourier transform of the dipole cross sections.

\end{document}